\documentstyle[12pt]{article}
\begin{document}
\title{Can wormholes exist?}
\author{V. Khatsymovsky \\
 {\em Institute of Theoretical Physics} \\
 {\em Box 803} \\
 {\em S-751 08 Uppsala, Sweden\thanks{Permanent adress (after 15
November 1993): Budker Institute of Nuclear Physics, Novosibirsk
630090, Russia}} \\
 {\em E-mail address: khatsym@rhea.teorfys.uu.se\thanks{Permanent
E-mail address (after 15 November): khatsym@inp.nsk.su}}}
\date{\setlength{\unitlength}{\baselineskip}
\begin{picture}(0,0)(0,0)
\put(9,13){\makebox(0,0){UUITP-20/1993}}
\put(8,12){\makebox(0,0){}}
\end{picture}
}
\maketitle
\begin{abstract}
Renormalized vacuum expectation values of electromagnetic
stress-energy tensor are calculated in the background
spherically-symmetrical metric of the wormhole's topology. Covariant
geodesic point separation method of regularization is used. Violation
of the weak energy condition at the throat of wormhole takes place
for geometry sufficiently close to that of infinitely long wormhole
of constant radius irrespectively of the detailed form of metric.
This is an argument in favour of possibility of existence of
selfconsistent wormhole in empty space maintained by vacuum field
fluctuations in the wormhole's background.
\end{abstract}
\newpage
{\bf 1.Introduction.} In 1988 Kip Thorne with co-workers
\cite{Thorne1},\cite{Thorne2} have
studied properties of static spherically-symmetrical traversable
wormhole. To exist, such the wormhole should be threaded by material
with rather unusual properties. In particular, radial pressure of the
material should exceed it's density both locally at the throat
\cite{Thorne1} and integrally along radial direction \cite{Thorne2}.
That is, weak energy condition (WEC) \cite{Haw} at the throat and
averaged weak energy condition (AWEC) \cite{TiRo} should be violated.
Where can we find such the material? In \cite{Thorne2} Casimir vacuum
between conducting spherical plates surrounding the throat was
considered as a kind of such the material since it possesses required
properties.
In the given note another possibility is studied: self-maintained
wormhole threaded by electromagnetic field vacuum. Covariantly
renormalized vacuum expectation values (VEV's) of electromagnetic
stress-energy tensor are found to violate  WEC at the throat of the
wormhole if it's typical length (proper radial distance at which
variation of radius becomes comparable to the radius itself) is large
as compared with it's radius. As for AWEC, check of it requires more
complicated calculation and will be given in forthcoming paper.

\bigskip
{\bf 2.Calculation.} In calculation it is convenient to use instead
of radius $ r$ the radial distance $ \rho$, so interval will take the
form

\begin{equation}
ds^2=r^2(\rho)d\Omega^2+d\rho^2-\exp{(2\Phi)}dt^2,~~~d\Omega^2=d\theta
^2+\sin^2{\!\theta}\,d\phi^2
\end{equation}
In the case of wormhole $ \Phi$ is everywhere finite \cite{Thorne1}.
Consider electromagnetic field in this metric background,
$F_{\mu\nu}=\partial_
\mu A_\nu -\partial_\nu A_\mu$, in the gauge $A_t =0$. There exist
transversal electric TE and transversal magnetic TM modes with
components of vector potential of the form

\begin{eqnarray}
{\rm TE:}~
(A_\rho,A_\theta,A_\phi)&=&(0,~\frac{Y_\phi}{\sin\theta}\;,~-Y_\theta\
sin\theta)R\\  {\rm TM:}~
(A_\rho,A_\theta,A_\phi)&=&(l(l+1)\frac{R}{r^2}Y,~R_\rho
Y_\theta,~R_\rho Y_\phi)\frac{\exp{(\Phi)}}{\omega}
\end{eqnarray}
Here $Y\equiv Y_{lm}(\theta,\phi)$,~$R\equiv
R_{nl}(\rho)$,~$\omega\equiv\omega_{nl}$,~$Y_\theta\equiv\partial
Y/\partial\theta$,~$Y_\phi\equiv\partial
Y/\partial\phi$,~$R_\rho\equiv dR/d\rho$.
Radial functions obey eigenvalue equation

\begin{equation}\label{rad.eq}
-\left(\exp{(\Phi)}{d \over d\rho} \right) ^2R+l(l+1){\exp{(2\Phi)}
\over r^2}R=\omega^2R
\end{equation}
with boundary conditions on conducting spherical plates $\Gamma$ (if
any)

\begin{equation}
{\rm TE}: ~R|_\Gamma=0;~~~{\rm TM}:~R_\rho|_\Gamma=0.
\end{equation}
It is apparently conformally invariant.

Substitute appropriately normalized solutions for it into
split-regularized form of  stress-energy tensor

\begin{equation}
T^{\mu\nu}(x)=\lim_{\tilde{x}\rightarrow
x}{[g^{\mu\lambda}(x)g^{\nu\omega}(x)-{1 \over
4}g^{\mu\nu}(x)g^{\lambda\omega}(x)]{1 \over
2}g^{\varrho\tau}(x)[F_{\lambda\varrho}(x)F_{\omega\tau}(\tilde{x})_x+
F_{\lambda\varrho}(\tilde{x})_xF_{\omega\tau}(x)]}
\end{equation}
and sum over eigenmodes (over $n,~l,~m$ and polarization $\sigma=$ TE
or TM). Here notation $F_{\mu\nu}(\tilde{x})_x$ means components of
tensor $F_{\mu\nu}(\tilde{x})$ transported in parallel way from point
$\tilde{x}$ to $x$ along the geodesic connecting both points. It is
convenient to split the point only in coordinate  $\rho$; then
parallel transport defined by vanishing the covariant derivative
means simply constancy of "physical" components,

\begin{equation}
A_{\hat{\mu}\hat{\nu}...\hat{\lambda}}\stackrel{\rm
def}{=}|g_{\mu\mu}g_{\nu\nu}...g_{\lambda\lambda}|^{-1/2}A_{\mu\nu...\
lambda}={\rm const},
\end{equation}

of any transported in such the way  tensor A. Notations coincide with
those of ref.\cite{Thorne1} and are convenient in what follows.
Normalization of radial functions should enable contribution of each
eigenmode to energy

\begin{equation}
-\int{T^\mu_{~t}g^{1/2}dS_\mu}=-\int{T^t_{~t}g^{1/2}d\rho d\theta
d\phi}~~~(g\equiv -{\rm det}\|g_{\mu\nu}\|)
\end{equation}
to take on vacuum value $ \hbar\omega/2$ (this is just {\it
covariant} component of  vector, as it should be for energy quantum
mechanically conjugated to {\it contravariant} time). By properties
of spherical functions summation over $ m$ eliminates angle
dependence. The resulting vacuum expectation values of components of
stress-energy tensor will be denoted just as components themselves;
this will not lead to any confusion. So we have for nonzero
components

\begin{eqnarray}
T_{\hat{t}\hat{t}}&=&\lim_{\epsilon\rightarrow 0}{\sum_{n,l,{\rm
TE,TM}}{{l+1/2 \over 8\pi\omega r\tilde{r}}{{R_\rho
\tilde{R}_\rho+\omega^2R\tilde{R}\exp{(-\Phi-\Phi^\prime)}+R\tilde{R}{
l(l+1)\over r\tilde{r}}} \over
\int{R^2\exp{(-\Phi)}d\rho}}}}\nonumber\\
T_{\hat{\rho}\hat{\rho}}&=&\lim_{\epsilon\rightarrow
0}{\sum_{n,l,{\rm TE,TM}}{{l+1/2 \over 8\pi\omega r\tilde{r}}{{R_\rho
\tilde{R}_\rho+\omega^2R\tilde{R}\exp{(-\Phi-\Phi^\prime)}-R\tilde{R}{
l(l+1)\over r\tilde{r}}} \over \int{R^2\exp{(-\Phi)}d\rho}}}}\\
T_{\hat{\theta}\hat{\theta}}&=&\lim_{\epsilon\rightarrow
0}{\sum_{n,l,{\rm TE,TM}}{{l+1/2 \over 8\pi\omega
r\tilde{r}}{{R\tilde{R}{l(l+1)\over r\tilde{r}}} \over
\int{R^2\exp{(-\Phi)}d\rho}}}}\nonumber
\end{eqnarray}
Here $ \epsilon=|\tilde{\rho}-\rho|$,~~$\tilde{R}\equiv
R(\tilde{\rho})$,~~$\tilde{r}\equiv r(\tilde{\rho})$~~and hereafter
the components of any tensor obtained by interchange
$\hat{\theta}\leftrightarrow\hat{\phi}$ are the same and will not be
written out.

Due to formal conformal invariance of electromagnetic field it is
natural to temporarily change variable $ \rho\rightarrow z$ via $
dz=\exp{(-\Phi)}d\rho$ and to consider the sums

\begin{eqnarray}
\sum_{l}^{}{(l+{1 \over 2})\sum_{n}^{}{{R_z\tilde{R}_z \over
\omega}}}&=&S_{\rm r}~(\equiv S_{\rm radial})\nonumber\\
\sum_{l}^{}{(l+{1 \over 2})\sum_{n}^{}{\omega R\tilde{R}}}&=&S_{\rm
f}~(\equiv S_{\rm full})\nonumber\\
\sum_{l}^{}{(l+{1 \over 2})l(l+1)\sum_{n}^{}{{R\tilde{R} \over
\omega}}}&=&S_{\rm a}~(\equiv S_{\rm angle})\\
\sum_{l}^{}{(l+{1 \over 2})\sum_{n}^{}{{R\tilde{R} \over
\omega}}}&=&U\nonumber
\end{eqnarray}
Then $ S_{\rm r}=U_{z\tilde{z}}$ and, by eq.(\ref{rad.eq}) $, S_{\rm
f}=-U_{z^2}+r^{-2}\exp{(2\Phi)}S_{\rm a}$ so that

\begin{eqnarray}\label{T-mu-nu}
8\pi T_{\hat{t}\hat{t}}&=&S_{\rm a}{\exp{(-\Phi)} \over
r^2\tilde{r}}\left( {\exp{\Phi} \over r}+{\exp{(\tilde{\Phi})} \over
\tilde{r}} \right)+{\exp{(-\Phi-\tilde{\Phi})} \over
r\tilde{r}}U_{z(\tilde{z}-z)}\nonumber\\
8\pi T_{\hat{\rho}\hat{\rho}}&=&S_{\rm a}{\exp{(-\Phi)} \over
r^2\tilde{r}}\left( {\exp{(\Phi)} \over r}-{\exp{(\tilde{\Phi})}
\over \tilde{r}} \right)+{\exp{(-\Phi-\tilde{\Phi})} \over
r\tilde{r}}U_{z(\tilde{z}-z)}\\
8\pi T_{\hat{\theta}\hat{\theta}}&=&S_{\rm a}{1 \over
r^2\tilde{r}^2}\nonumber
\end{eqnarray}

Computation of sums over $ n$ in the above expressions reduces to
that  of a Green function by the following trick. We introduce

\begin{equation}
G^{\rm E~or~M}\equiv G^{\rm E~or
{}~M}_l(t,\rho,\tilde{\rho})=\sum_{n,{\rm TE~or~ TM}}^{}{{R\tilde{R}
\over t^2+\omega^2}},~~~G=G^{\rm E}+G^{\rm M}
\end{equation}
through which $ U$ and $ S_{\rm a}$ are expressible:

\begin{eqnarray}\label{USa}
U&=&\sum_{l=1}^{\infty}{\int_{0}^{\infty}{{2l+1 \over
\pi}Gdt}}\equiv\sum{}\!\!\!\!\!\!\!\!\int{{2l+1 \over
\pi}Gdt}\nonumber\\
S_{\rm a}&=&\sum{}\!\!\!\!\!\!\!\!\int{{2l+1 \over \pi}l(l+1)Gdt}
\end{eqnarray}
In turn,  $ G^{\rm E},~G^{\rm M}$ are some Green functions:

\begin{eqnarray}
-G^{\rm E,M}_{zz}+\left(t^2+l(l+1){\exp{(2\Phi)} \over
r^2}\right)G^{\rm E,M}=\delta(z-z^\prime)\\
G^{\rm E}|_{z\in\Gamma}=0,~~~G^{\rm
M}_z|_{z\in\Gamma}=0~~~~~~~~~~~~~~~~~\nonumber
\end{eqnarray}
These eqs. are then solved perturbatively. Put

\begin{equation}
{\exp{(2\Phi)} \over r^2}={\exp{(2\Phi_0)} \over r^2_0}+V,\nonumber
\end{equation}
where index $ 0$ denote values of considered quantities at  the
throat (at $ \rho=0$), and expand over $ V$.

We are interested here in empty space and thus shift $ \Gamma$ to
infinity; then it proves that $ G^{\rm E}=G^{\rm M}\equiv
\bar{G},~G=2\bar{G}$. In zero approximation ($ V=0$)

\begin{equation}\label{G}
\bar{G}^{(0)}={1 \over 2m}\exp{(-m|z-\tilde{z}|)},~~~m^2\equiv
t^2+l(l+1){\exp{(2\Phi_0)} \over r^2_0}.
\end{equation}
In the first  (linear in $ V$) approximation

\begin{equation}\label{pert}
G=2\bar{G}^{(0)}(z,\tilde{z})-2l(l+1)\int{\bar{G}^{(0)}(z,y)V(y)\bar{G
}^{(0)}(y,\tilde{z})dy}
\end{equation}

Calculating so, we are faced with integrosums over $ l,~dt$. There
are two basic ones, and these arise already in zero approximation. In
this approximation, substituting variables $
t=\exp{(\Phi_0)}r^{-1}_0[l(l+1)]^{1/2}\sinh{\varphi}$ and then $
\epsilon\cosh{\varphi}=r_0q$ we get with the help of (\ref{G}),
(\ref{USa}) and (\ref{T-mu-nu})

\begin{equation}
2\pi^2 r^4_0\left (\matrix{
T_{\hat{t}\hat{t}}\cr
T_{\hat{\rho}\hat{\rho}}\cr
T_{\hat{\theta}\hat{\theta}}\cr
}\right )=\left (\matrix{
-I_1(\epsilon/r_0)\cr
-I_1(\epsilon/r_0)-I_2(\epsilon/r_0)\cr
I_2(\epsilon/r_0)/2\cr
}\right )
\end{equation}
where

\begin{equation}
I_1(\epsilon)={1 \over
\epsilon^2}\int_{\epsilon}^{\infty}{{(q^2-\epsilon^2)^{1/2} \over
q^4}h(q)dq},~~~I_2(\epsilon)=\int_{\epsilon}^{\infty}{{h(q)dq \over
q^4(q^2-\epsilon^2)^{1/2}}}
\end{equation}
and function $ h(q)$ is regular at $ q=0$. Carefully expanding it
under the integral sign in Taylor series we get

\begin{eqnarray}
\!\!\!I_1(\epsilon)&\!\!\!=&\!\!\!{1 \over 3}{h(0) \over
\epsilon^4}+{\pi \over 4}{h^\prime(0) \over \epsilon^3}+{1 \over
2\epsilon^2}\left[\int_{0}^{\infty}{{h^{\prime\prime}(q)-h^{\prime\prime}
(0)\theta(M-q)\over q}dq}+h^{\prime\prime}(0)\left( \ln{{2M \over
\epsilon}} +{1 \over 2}\right) \right]\nonumber\\&&\!\!\!-{\pi \over
12}{h^{\prime\prime\prime}(0) \over \epsilon}-{1 \over 48}\left[
\int_{0}^{\infty}{{h^{\imath {\rm v}}(q)-h^{\imath {\rm
v}}(0)\theta(M-q)\over q}dq}+h^{\imath {\rm v}}(0)\left( \ln{{2M
\over \epsilon}} +{31 \over 12}\right)\right],\nonumber\\
\!\!\!I_2(\epsilon)&\!\!\!=&\!\!\!{2 \over 3}{h(0) \over
\epsilon^4}+{\pi \over 4}{h^\prime(0) \over \epsilon^3}+{1 \over
2}{h^{\prime\prime}(0) \over \epsilon^2}+{\pi \over
12}{h^{\prime\prime\prime}(0) \over \epsilon}\\&&\!\!\!+{1 \over
24}\left[ \int_{0}^{\infty}{{h^{\imath {\rm v}}(q)-h^{\imath {\rm
v}}(0)\theta(M-q)\over q}dq}+h^{\imath {\rm v}}(0)\left( \ln{{2M
\over \epsilon}} +{25 \over 12}\right)\right]\nonumber
\end{eqnarray}
For $ h(q)$ we have

\begin{eqnarray}
h(q)&=&q^4\sum_{l=1}^{\infty}{\left(l+{1 \over
2}\right)l(l+1)\exp{\{-q[l(l+1)]^{1/2}\}}}\nonumber\\
&=&q^4{d^2 \over dq^2}{f(q) \over
q^2}=q^2f^{\prime\prime}(q)-4qf^\prime(q)+6f(q),\\
f(q)&=&q^2\sum_{l=1}^{\infty}{\left(l+{1 \over
2}\right)\exp{\{-q[l(l+1)]^{1/2}\}}}\nonumber
\end{eqnarray}
The $ f(q)$ is also regular at $ q=0$. Thus $ h(0)=6f(0)$, $
h^\prime(0)=2f^\prime(0)$, $ h^{\prime\prime}(q)=q^2f^{\imath {\rm
v}}(q)$, $ h^{\prime\prime\prime}(0)=0$, $ h^{\imath {\rm
v}}(0)=2f^{\imath {\rm v}}(0)$. In $ f(q)$ we expand  exponential
denoting $ l+{1 \over 2}=\gamma$:

\begin{eqnarray}
&&\exp{\left[ -q\left( \gamma^2-{1 \over 4}
\right)^{1/2}\right]}=\exp{(-q\gamma)}\left[ 1+qF+{1 \over
2}q^2F^2+O(q^3F^3)\right],\nonumber\\
&&F=\gamma\left[1-\left( 1-{1 \over 4\gamma^2} \right)^{1/2}
\right]={1 \over 8\gamma}+{1 \over 128\gamma^3}+O(\gamma^{-5})
\end{eqnarray}
This is enough to express required derivatives of $ f(q)$ at zero in
terms of those of elementary sum $ g(q)=q\sum{\exp{(-q\gamma)}}$.
Lengthy also elementary calculation ultimately gives

\begin{equation}\label{T-zero}
2\pi^2 r^4_0\left (\matrix{
T_{\hat{t}\hat{t}}\cr
T_{\hat{\rho}\hat{\rho}}\cr
T_{\hat{\theta}\hat{\theta}}\cr
}\right )^{(0)}_{\rm reg}=\left (\matrix{
-2{r^4_0 \over \epsilon^4}+{1 \over 3}{r^2_0 \over \epsilon^2}-{1
\over 60}\ln{{L \over \epsilon}}~~~~~~~\cr
-6{r^4_0 \over \epsilon^4}+{1 \over 3}{r^2_0 \over \epsilon^2}+{1
\over 60}(\ln{{L \over \epsilon}}-1)\cr
+2{r^4_0 \over \epsilon^4}~~~~~~~~-{1 \over 60}(\ln{{L \over
\epsilon}}-{1 \over 2})\cr
}\right )
\end{equation}
for regularized stress-energy tensor in zero approximation (for
space-time taken as direct product of sphere and Minkowsky plain).
The nonlogarithmic $ O(h^{\imath {\rm v}})$ contribution in $
I_1,~I_2$ cannot be accurately estimated within our elementary
approach and is simply included into logarithm thus resulting in the
appearance of effective cut off $ L=kr_0$ with some numerical value $
k$ of the order of unity.

Now one renormalizes gravity action by adding to it (and subtracting
from matter action) some functional $ W_{\rm div}$ so that

\begin{equation}
T^{\mu\nu}_{\rm div}=g^{-1/2}{\delta W_{\rm div} \over \delta
g_{\mu\nu}}
\end{equation}
cancels divergences in matter $ T^{\mu\nu}$ \cite{deWitt}. This
results in renormalization of cosmological constant, Einstein gravity
constant and Weyl tensor invariant term in the action,

\begin{equation}
\int{C_{\mu\nu\lambda\omega}C^{\mu\nu\lambda\omega}g^{1/2}d^4x},
\end{equation}
 where $C_{\mu\nu\lambda\omega} $ is conformal Weyl tensor. The full
expression for $ T^{\mu\nu}_{\rm div}$ was given by Christensen
\cite{Christ}. Nontrivial here is appearance of finite part of $
T^{\mu\nu}_{\rm div}$ leading to nonzero trace of renormalized
stress-energy tensor,

\begin{equation}
T^{\mu\nu}_{\rm ren}=T^{\mu\nu}_{\rm reg}-T^{\mu\nu}_{\rm div},
\end{equation}
so that

\begin{equation}
T^\mu_{~\mu, {\rm ren}}={1 \over 2880\pi^2}\left(
-13R_{\mu\nu\lambda\omega}R^{\mu\nu\lambda\omega}+88R_{\mu\nu}R^{\mu\nu}
-25R^2-18{\,\lower0.9pt\vbox{\hrule \hbox{\vrule height 0.2 cm
\hskip 0.2 cm \vrule height 0.2 cm}\hrule}\,}R \right)
\end{equation}
Christensen's result depends locally on Riemannian and metric tensors
and on $ \epsilon^\mu$, the vector tangential at $ x$ to geodesic
connecting points $ x$ and $ x^\prime$, with length $ \epsilon$, the
length of geodesic.  In our case $ \epsilon^\mu=(0,~0,~\epsilon,~0)$.
In zero approximation (space-time$ =$sphere$\times$plain) the
Riemannian tensor has the only (up to index permutations) nonzero
component $
R_{\hat{\theta}\hat{\phi}\,\hat{\theta}\hat{\phi}}=r^{-2}_0$.
Substitute these input values into Christensen's formula and subtract
the resulting  $ T^{\mu\nu}_{\rm div}$ from our answer
(\ref{T-zero}) for $T^{\mu\nu}_{\rm reg}$. All the divergent terms,$
\epsilon^{-4},~\epsilon^{-2}$ and $ \ln{\epsilon}$ ones thus get
cancelled. On the contrary, their cancellation is a useful check of
our calculation. In logarithms the UV regulator $ \epsilon$ is
substituted by infrared one $ \Lambda$. The finite part of $
T^{\mu\nu}_{\rm div}$ turns out to be nonzero in this simple case of
geometry only for one, $ \rho\rho$-component:

\begin{equation}
2\pi^2 r^4_0\left (\matrix{
T_{\hat{t}\hat{t}}\cr
T_{\hat{\rho}\hat{\rho}}\cr
T_{\hat{\theta}\hat{\theta}}\cr
}\right )^{(0)}_{\rm div, finite}=\left (\matrix{
{}~~0\cr
-{1 \over 60}\cr
{}~~0\cr
}\right ),
\end{equation}
i.e. it is completely consists of anomaly. The resulting renormalized
tensor takes the form

\begin{equation}
2\pi^2 r^4_0\left (\matrix{
T_{\hat{t}\hat{t}}\cr
T_{\hat{\rho}\hat{\rho}}\cr
T_{\hat{\theta}\hat{\theta}}\cr
}\right )^{(0)}_{\rm ren}=\left (\matrix{
{1 \over 60}\ln{{\Lambda \over L}}~~~~~~~\cr
-{1 \over 60}\ln{{\Lambda \over L}}~~~~~~~~~~\cr
{1 \over 60}\ln{{\Lambda \over L}}+{1 \over 120}\cr
}\right ).
\end{equation}
The $ \rho - t$ symmetry inherent in this simple case is restored
just due to anomaly.

We now consider deviation of $ \Phi, r$ from constants. Since we
cannot take into account such deviation precisely, we need some
parameter to expand over it. This parameter may be the ratio of
wormhole's radius $ r(0)$ to the typical length - the proper radial
distance $ \rho$ at which the variation $ r(\rho)-r(0)$ is comparable
to $ r(0)$. In practice, it is equivalent to expansion over the full
number of differentiations $ {d \over d\rho}$. For example, $
\Phi^{\prime\prime}$ and $ \Phi^{\prime 2}$ are considered in such
the expansion on equal footing. At the throat odd derivatives vanish
due to symmetry, and the nearest correction to zero approximation
will be given by $ \Phi^{\prime\prime}$-,  $ r^{\prime\prime}$-terms
arising in linear approximation over deviation of $ \Phi,~r$ from
constants. This will lead to the following three types of
contributions to $ T^{\mu\nu}_{\rm reg}$ at the throat ($ \rho=0$).

First, those due to dependence on $ \epsilon$ of metric factors in
(\ref{T-mu-nu}) taken at $ \tilde{x}$; these follow by expanding in
Taylor series in $ \epsilon$. At the throat we have:

\begin{eqnarray}
\exp{(\tilde{\Phi})}&=&\exp{(\Phi_0)}\left[ 1+{1 \over
2}\Phi_0^{\prime\prime}\epsilon^2+O(\epsilon^4) \right],\nonumber\\
\tilde{r}&=&r_0+{1 \over 2}r_0^{\prime\prime}\epsilon^2+O(\epsilon^4)
\end{eqnarray}
(here $ r_0^{\prime\prime}\equiv r^{\prime\prime}(0)$ etc.).

Second, those connected with changing expression of $
|\tilde{z}-z|=|\int_{0}^{\epsilon}{\exp{(-\Phi)}d\rho}|$ entering
already found formulas for $ T^{\mu\nu\>(0)}_{\rm reg}$ in terms of
geodesic length $ \epsilon$; to take this changing into account one
should substitute in these formulas $ \epsilon$ by

\begin{eqnarray}
\epsilon\left[ 1-{1 \over
6}\Phi_0^{\prime\prime}\epsilon^2+O(\epsilon^4) \right]\nonumber
\end{eqnarray}

Third, contribution coming from linear in $ V$ correction to Green
function $ G$. Here we simply substitute Taylor series for $ V$
around $ y=0$ into perturbative expansion (\ref{pert}) for $ G$ and
integrate over $ y$; due to exponents integration for each Taylor
term is finite. Appearing integrosums reduce to that of above
considered type $ I_2$, but now  $ q^{-4}h$ is less singular at $
q=0$ and/or it gets multiplied by positive power of $ \epsilon$.

These contributions are collected in the resulting  total correction
$ T^{\mu\nu\>(1)}_{\rm reg}$, respectively:

\begin{eqnarray}
2\pi^2\left (\matrix{
T_{\hat{t}\hat{t}}\cr
T_{\hat{\rho}\hat{\rho}}\cr
T_{\hat{\theta}\hat{\theta}}\cr
}\right )^{(1)}_{\rm reg}&=&\left (\matrix{
{}~~~~~~~~~~~~2{\Phi_0^{\prime\prime} \over \epsilon^2}-{1 \over
6}{r_0^{\prime\prime} \over r_0^3}-{1 \over 6}{\Phi_0^{\prime\prime}
\over r_0^2}\cr
{}~4{r_0^{\prime\prime}/r_0 \over \epsilon^2}+2{\Phi_0^{\prime\prime}
\over \epsilon^2}-{1 \over 6}{r_0^{\prime\prime} \over r_0^3}-{1
\over 6}{\Phi_0^{\prime\prime} \over r_0^2}\cr
-2{r_0^{\prime\prime}/r_0 \over
\epsilon^2}~~~~~~~~~~~~~~~~~~~~~~~~~~\cr
}\right )+\left (\matrix{
-{4 \over 3}{\Phi_0^{\prime\prime} \over \epsilon^2}+{1 \over
9}{\Phi_0^{\prime\prime} \over r_0^2}\cr
-4{\Phi_0^{\prime\prime} \over \epsilon^2}+{1 \over
9}{\Phi_0^{\prime\prime} \over r_0^2}\cr
{}~{4 \over 3}{\Phi_0^{\prime\prime} \over \epsilon^2}~~~~~~~~\cr
}\right )\nonumber\\&+&\left (\matrix{
{}~{2 \over 3}{r_0^{\prime\prime}/r_0 \over \epsilon^2}-{2 \over
3}{\Phi_0^{\prime\prime} \over \epsilon^2}-{1 \over
18}{r_0^{\prime\prime} \over r_0^3}+{1 \over
18}{\Phi_0^{\prime\prime} \over r_0^2}\cr
-{8 \over 3}{r_0^{\prime\prime}/r_0 \over \epsilon^2}+{8 \over
3}{\Phi_0^{\prime\prime} \over \epsilon^2}~~~~~~~~~~~~~~~~~~~~\cr
+{5 \over 3}{r_0^{\prime\prime}/r_0 \over \epsilon^2}-{5 \over
3}{\Phi_0^{\prime\prime} \over \epsilon^2}-{1 \over
36}{r_0^{\prime\prime} \over r_0^3}+{1 \over
36}{\Phi_0^{\prime\prime} \over r_0^2}\cr
}\right ).
\end{eqnarray}

With the same accuracy we get for components of Riemannian tensor at
the throat (up to index permutations and $
\hat{\theta}\leftrightarrow\hat{\phi}$):

\begin{equation}
R_{\hat{\theta}\hat{\phi}\,\hat{\theta}\hat{\phi}}=r^{-2}_0,~~~R_{\hat
{\rho}\hat{\theta}\hat{\rho}\hat{\theta}}=-{r^{\prime\prime}_0 \over
r_0},~~~R_{\hat{t}\hat{\rho}\hat{t}\hat{\rho}}=\Phi^{\prime\prime}_0.
\end{equation}
Others are zero in this approximation. Using Christensen's formula
for $ T^{\mu\nu}_{\rm div}$ with this input and subtracting it from $
T^{\mu\nu}_{\rm reg}$ cancels all the singularities, as it should;
besides that, we have for correction to finite part:

\begin{equation}
2\pi^2\left (\matrix{
T_{\hat{t}\hat{t}}\cr
T_{\hat{\rho}\hat{\rho}}\cr
T_{\hat{\theta}\hat{\theta}}\cr
}\right )^{(1)}_{\rm div,finite}=\left (\matrix{
-{1 \over 12}{r_0^{\prime\prime} \over r_0^3}~~~~~~~~~~\cr
-{1 \over 6}{r_0^{\prime\prime} \over r_0^3}-{1 \over
18}{\Phi_0^{\prime\prime} \over r_0^2}\cr
-{1 \over 12}{r_0^{\prime\prime} \over r_0^3}-{13 \over
72}{\Phi_0^{\prime\prime} \over r_0^2}\cr
}\right )
\end{equation}
With taking into account all the contributions we ultimately get for
total renormalized tensor in the considered approximation:

\begin{equation}
2\pi^2 \left (\matrix{
T_{\hat{t}\hat{t}}\cr
T_{\hat{\rho}\hat{\rho}}\cr
T_{\hat{\theta}\hat{\theta}}\cr
}\right )^{(0)+(1)}_{\rm ren}=\left (\matrix{
{1 \over 60r^4_0}\ln{{\Lambda \over L}}~~~~~~~~-{5 \over
36}{r_0^{\prime\prime} \over r_0^3}~~~~~~~~~~\cr
-{1 \over 60r^4_0}\ln{{\Lambda \over
L}}~~~~~~~~~~~~~~~~~~~~~~~~~~~\cr
{1 \over 60r^4_0}\left(  \ln{{\Lambda \over L}}+{1 \over 2}\right)-{1
\over 36}{r_0^{\prime\prime} \over r_0^3}+{5 \over
72}{\Phi_0^{\prime\prime} \over r_0^2}\cr
}\right ).
\end{equation}

\bigskip
{\bf 3.Discussion.} Taking the difference between radial pressure $
\tau=-T_{\hat{\rho}\hat{\rho}}$ and energy density $
\varrho=T_{\hat{t}\hat{t}}$ obtained we find it to be positive at the
throat:

\begin{equation}
\tau-\varrho={5 \over 72\pi^2}{r^{\prime\prime}_0\over r^3_0}>0.
\end{equation}
This is quite nontrivial result, since generally we might get
arbitrary combination of $
\Phi^{\prime\prime}_0,~~r^{\prime\prime}_0/r_0$, not necessarily
simply $ r^{\prime\prime}_0/r_0$ with positive coefficient. Thus, WEC
is violated at the throat, at least on the level of second
derivatives: not only LHS of gravity eqs. (Einstein tensor) implies $
\tau>\varrho$, but also this condition is fulfilled by RHS
(electromagnetic stress-energy tensor) for given topology. As for the
derivatives of larger order,  their contribution to $ \tau-\varrho$
is not a'priori fixed in sign, but the fact that required sign is
achieved already on the level of second derivatives, means that
contribution of higher derivatives required for further adjustment of
given value need not be larger then that of first ones, which is
normal condition of applicability of perturbation theory and
regularity of possible solution.

Large logarithms do not contribute $ \tau-\varrho$ in the given
approximation. This is because corrections to the structure

\begin{equation}\label{CC}
g^{-1/2}{\delta  \over \delta
g_{\mu\nu}}\int{C_{\mu\nu\lambda\omega}C^{\mu\nu\lambda\omega}g^{1/2}d
^4x}
\end{equation}
entering $ T^{\mu\nu}$ as coefficient at logarithm, prove to occur
only at the level of four differentiations. To see effect of the
fourth derivatives we linearly expand (\ref{CC}). Einstein eqs. at
the throat read:

\begin{eqnarray}
R_{\hat{t}\hat{t}}-{1 \over 2}Rg_{\hat{t}\hat{t}}&=&{1 \over
r^2_0}-2{r^{\prime\prime}_0 \over r_0}={\kappa \over 2\pi^2}\left\{
{N \over 60r^4_0}\left[ 1+2r^4_0\left( \Phi^{\imath {\rm
v}}_0-{r^{\imath {\rm v}}_0\over r_0} \right) \right]-{5 \over
36}{r^{\prime\prime}_0 \over r^3_0} \right\},\nonumber\\
R_{\hat{\rho}\hat{\rho}}-{1 \over 2}Rg_{\hat{\rho}\hat{\rho}}&=&-{1
\over r^2_0}~~~~~~={\kappa \over 2\pi^2}\left( -{N \over 60r^4_0}
\right),\\
R_{\hat{\theta}\hat{\theta}}-{1 \over
2}Rg_{\hat{\theta}\hat{\theta}}&=&\Phi^{\prime\prime}_0+~~{r^{\prime\prime}_0
\over r_0}\nonumber\\
&=&{\kappa \over 2\pi^2}\left\{ {N \over 60r^4_0}\left[ 1+r^4_0\left(
\Phi^{\imath {\rm v}}_0-{r^{\imath {\rm v}}_0\over r_0} \right)
\right]+{1 \over 120r^4_0}
-{1 \over 36}{r^{\prime\prime}_0 \over r^3_0}+{5 \over
72}{\Phi^{\prime\prime}_0 \over r^2_0}\right\}.\nonumber
\end{eqnarray}
Here $ N=\ln{{\Lambda \over L}}$, $ \kappa=8\pi G$, $ G$ is Newtonian
gravity constant. Not shown are bilinear in second derivatives terms
multiplying $ N$ which in precise calculations should be taken into
account on equal footing with fourth derivatives. From $ \rho\rho$ -
equation the estimate for wormhole radius follows,

\begin{equation}\label{radius}
r^2_0={\kappa \over 240\pi^2}\ln{\left({240\pi^2 \over
\kappa}{\Lambda^2 \over k^2}\right)}
\end{equation}
(we have taken into account that $ N$ itself depends on $ r_0$
through $ L=kr_0$, $ k\sim 1$). Then it follows from two other eqs.
that

\begin{equation}
\Phi^{\prime\prime}_0, {r^{\prime\prime}_0 \over r_0}\sim {1 \over
r^2_0},~~~\Phi^{\imath {\rm v}}_0, {r^{\imath {\rm v}}_0\over
r_0}\sim {1 \over r^4_0}.
\end{equation}
In other words, the length of the wormhole is of the order of it's
radius. Thus, approximation of long wormhole seems to be not relevant
to selfconsistent solution, but we can hope to approach qualitatively
physical wormhole starting from the sufficiently long one.

As for the range of possible values of $ r_0$, it follows from the
experimental limitations on $ \Lambda$. The latter arise in
connection with that the induced $ C^2$-term should violate
experimentally observed planetary motion, especially that of mercury
\cite{deWitt}. The $ C^2$-term enters action with coefficient just of
the order of $ r^2_0 /\kappa$ (the $ r^2_0$ itself as given by
(\ref{radius}) is defined by this coefficient), i.e. with coefficient
$ \sim r^2_0$ relative to Einstein $ R$-term. So the effect of  $
C^2$-term has relative value like $ \exp{(-r_{\rm merc}/r_0)}$
\cite{deWitt}. Here $ r_{\rm merc}$ is typical radius of mercury
orbit. So the requirement for this correction not to exceed
experimental accuracy leads to $ r_0$ being at least 1.5 orders of
magnitude smaller than the radius of mercury orbit, i.e. of the order
of the radius of sun. As for the lower limit for $ r_0$, it can be
estimated by taking $ \Lambda\sim 10^{62}$ in Plank units, which
corresponds to experimentally observable part of Universe.

Thus, the range of experimentally acceptable (in indirect way) values
of $ r_0$ is quite large: it can be by half an order smaller than the
Plank length (though in this region the concept of classical gravity
may be wrong) and it can be as large as radius of sun.

Finally, these estimates do not take into account the influence of
another fields, of which the most essential may be that of massless
particles (e.g. the neutrino one) due to occurrence of large
logarithms. Corresponding work is in progress now.

\bigskip
I am grateful to prof. A. Niemi, S. Yngve and personnel of Institute
of Theoretical Physics at Uppsala University for warm hospitality and
support during the work on this paper.

\end{document}